\documentclass[10pt]{article}
\usepackage{amssymb,amsfonts,amsmath,latexsym}
\usepackage{graphics,graphicx,epsf}
\newcommand{\be}{\begin{equation}}
\newcommand{\ee}{\end{equation}}
\newcommand{\bea}{\begin{eqnarray}}
\newcommand{\eea}{\end{eqnarray}}
\newcommand{\ba}{\begin{array}}
\newcommand{\ea}{\end{array}}
\newcommand{\bt}{\begin{tabular}}
\newcommand{\et}{\end{tabular}}

\newcommand{\fr}{\frac}
\newcommand{\ci}{\cite}
\newcommand{\cl}{\centerline}
\newcommand{\bs}{\bigskip}

\newcommand{\vs}{\vspace}

\newcommand{\en}{\eqno}

\newcommand{\bbib}{\begin{thebibliography}}
\newcommand{\ebib}{\end{thebibliography}}

\newcommand{\lra}{\leftrightarrow}

\begin{document}

\cl{\bf ON UNIVERSALITY OF CONDUCTIVITY}
\cl{\bf OF PLANAR  RANDOM SELF-DUAL SYSTEMS}

\bs

\cl{\bf S.A.Bulgadaev \footnote{e-mail: bulgad@itp.ac.ru}}

\bs
\cl{Landau Institute for Theoretical Physics}
\cl{Chernogolovka, Moscow Region, Russia, 142432}

\bs

\begin{quote}
\footnotesize{
General properties of the effective conductivity
$\sigma_e$ of planar isotropic randomly inhomogeneous
 two-phase self-dual systems are investigated.
A new approach for finding out $\sigma_e$ of random  systems
based on a duality,  a series
expansion in the inhomogeneous parameter $z$ and additional assumptions,
is proposed. Two new approximate expressions for $\sigma_e$
at arbitrary values of phase concentrations are found.
They satisfy all necessary inequalities, symmetries, including a dual
one, and reproduce known results in various limiting cases.
Two corresponding models with different inhomogeneity structures, whose
$\sigma_e$ coincide with these  expressions, are
constructed. First model describes systems with a finite maximal characteristic
scale of the inhomogeneities. In this model $\sigma_e$ is a solution of
the approximate functional equation, generalizing the duality
relation. The second model is constructed from squares with random layered
structure.
The difference of $\sigma_e$ for these models means a nonuniversality of the
effective conductivity even for binary random self-dual systems.
The first explicit expression for $\sigma_e$ can be used also for approximate
description of various inhomogeneous systems with compact inclusions
of the second phase.
The percolation problem of these models is briefly discussed.
}
\end{quote}

\bs

\bs

\cl{\bf 1. Introduction}
\bs

The electrical transport properties of the classical inhomogeneous condensed
media, being a particular case of the transport properties in disordered systems,
have a great practical interest. For this reason they have been intensively
studied theoretically as well as experimentally (see, for example, \ci{1,2}).
An importance of this knowledge has revived last time. It was
established that many real systems, which have been
considered to be homogeneous, can have  on small scales (from microscopic
through mesoscopic till macroscopic) inhomogeneous or even heterophase
structures.
In particular, such situation takes place in the oxide materials, which are
now actively studied \ci{3}. Analogous situation takes place in
many real examples of the disordered systems (for example \ci{4}).
For this reason the observed effective conductivity $\sigma_{e}$ of these
systems can differ from that of homogeneous media. Thus one needs a theory
adequately describing conducting properties of such inhomogeneous media.
Unfortunately, the theory of strongly inhomogeneous systems is not well
developed due to a difficulty of explicit averaging over strong inhomogeneous
fluctuations. The various perturbation, approximate and variation theories
have been developed and many interesting  results
have been obtained so far (see, for example, \ci{5} and a recent review \ci{6}).

In the conducting properties of the inhomogeneous media there is one old
problem about the effective
conductivity $\sigma_{e}$ of inhomogeneous (regular, irregular or random)
heterophase systems, containing $N (N \ge 2)$ various phases with
different partial conductivities  $\sigma_i  \; (i =1,2,...,N).$
In general, the inhomogeneities can be very different, changing from
regular through  weakly and moderately irregular till strongly
irregular ones. Strong inhomogeneous fluctuations can be connected with
large differences of the partial conductivities as well as with
strong irregularities of these phases. The latter situation is very similar
to the  case of random inhomogeneous medium and in some cases can be
identified with it.

In principle, the regular and weakly irregular cases can be
considered systematically in framework of classical potential theory \ci{7},
while an investigation of strong and
random inhomogeneous systems needs some  approaches connected with
the averaging over the random regions with different conductivities \ci{8}.

The basic equation, describing a stationary electric current in random systems,
is the Ohm law, connecting the local current
${\bf j}({\bf r})$ and the local electric field ${\bf E}({\bf r}),$
$$
{\bf j}({\bf r}) = \sigma ({\bf r}) \;{\bf E}({\bf r}),
\en(1)
$$
where $\sigma ({\bf r})$ is a local conductivity. We will assume that the
carriers in all phases have the charges of the same sign. Then one  can
suppose that $\sigma ({\bf r}) \ge 0.$ The equation (1) must be
supplemented with the corresponding boundary conditions on the boundaries of
two phases \ci{1}
$$
j^1_n = j^2_n, \quad E^1_t = E^2_t,
\en(2)
$$
where $n,t$ denote the normal and tangent components and $1,2$ correspond
to different phases.
The vector fields ${\bf E}({\bf r})$ and ${\bf j}({\bf r})$ satisfy the
following equations inside each phase
$$
\nabla \times {\bf E}({\bf r}) = 0, \quad
\nabla \cdot {\bf j}({\bf r}) = 0.
\en(3)
$$
They mean that ${\bf E}({\bf r})$  and
${\bf j}({\bf r})$ are  potential and  divergenceless fields respectively.

The effective conductivity $\sigma_{e}$ of the isotropic randomly and/or
strongly  inhomogeneous medium can be defined as
a proportionality coefficient between mean  values of ${\bf j}$ and ${\bf E}$
averaged over the area of the system
$$
\bar {\bf j} = \sigma_{e} \bar {\bf E},
\en(4)
$$
$$
\bar {\bf j} \equiv \int  {\bf j}({\bf r}) d{\bf r}/S, \quad
\bar {\bf E} \equiv \int  {\bf E}({\bf r}) d{\bf r}/S
\en(5)
$$
where $S$ is an area of the system.

Since it is very difficult to make an averaging of (4,5)  over an area
of the system explicitly for strongly irregular or random systems
it is usually supposed that for systems with such
$\sigma ({\bf r})$ (or with a fixed random realization of $\sigma ({\bf r})$
the averaging of (4,5)  over the area of the system can be replaced by the
averaging   over some
ensemble of random functions $\sigma ({\bf r})$ with a functional weight
$\cal P\{\sigma\},$ which may, in general, depend also on temperature T
$$
\int d {\bf r} F(\sigma ({\bf r}))/S = \int {\cal D}\sigma({\bf r})
F(\sigma ({\bf r})) {\cal P}\{\sigma\} \equiv \;\ll F(\sigma ({\bf r})) \gg.
\en(6)
$$
In this paper we confine ourselves with an investigation of the $\sigma_e$
dependence on  $\sigma_i$ and assume that T is fixed.

According to the general principles, a behaviour of the effective conductivity
of random systems must depend on statistical properties of the random
conductivity $\sigma ({\bf r})$ encoded in its correlations. In case of random
heterophase systems one usually supposes that $\sigma ({\bf r})$
at different space points are statistically independent and its moments
 in each space point ${\bf r}$ are defined by one
distribution function of partial conductivities $P(\sigma),$
which does not depend on ${\bf r},$
$$
P(\sigma) = \sum_{i=1}^N x_i \delta (\sigma - \sigma_i), \quad
\sum_1^N x_i =1, \quad
\int P(\sigma) d\sigma = 1,
\en(7)
$$
where $x_i$ is the concentration of the i-th
phase.

This means that a mean conductivity $\ll \sigma \gg$ (and all its higher moments)
can be represented in two equivalent forms
$$
\ll \sigma \gg = \int \sigma ({\bf r}) d {\bf r}/S =
\int \sigma P(\sigma) d\sigma = \langle \sigma \rangle.
\en(8)
$$
where $\langle \ldots \rangle$ denotes the
average with a distribution function $P(\sigma)$
Then the bare phenomenological correlation function of conductivities $C_0(r)$
takes the simple form
$$
C_0(r,r') =  \ll \sigma (r) \sigma (r') \gg  =
\langle \sigma \rangle^2 + \langle \sigma^2 \rangle \delta (r-r')
\en(9)
$$
As follows from (9) $C_0(r,r')$ in this form does not already
describe a detailed spatial structure of random inhomogeneities,
but contains only their integral characteristics -- the concentrations
of phases. In this case $\sigma_e$ can be represented as a function of partial
conductivities $\sigma_i$ and concentrations $x_i$ only
$$
\sigma_e = \sigma_e(\sigma_1,x_1|\sigma_2,x_2|...|\sigma_N,x_N)
\en(10)
$$
Due to the linearity of the defining equations this function must be a
homogeneous function of degree one
of $\sigma_i, i=1,...,N.$ It has also satisfy the inequalities following from
general principles (see, for example, \ci{6})
$$
\langle \sigma^{-1} \rangle^{-1} \le \sigma_e \le \langle \sigma \rangle.
\en(11)
$$
For practical purposes and for a description of experimental data
one needs usually some explicit and compact (though, maybe, approximate)
formulas for effective conductivity applicable in a wide range of parameters.
A general expression for $\sigma_{e}$ of isotropic media at arbitrary
concentrations $x_i$ can be obtained only
in case of weakly inhomogeneous media, when the conductivity
fluctuations $\Delta \sigma$
($\Delta \sigma = \sigma - \langle \sigma \rangle$,
are much smaller than a mean conductivity $\langle \sigma \rangle$ \ci{1}
$$
\sigma_{e} = \langle \sigma \rangle \left(1 -
\fr{\langle (\delta \sigma)^2 \rangle}
{D}\right), \quad  \delta \sigma = \frac{\Delta \sigma}{\langle \sigma \rangle},
\quad \langle (\Delta \sigma)^2 \rangle =
\langle \sigma^2 \rangle - {\langle \sigma \rangle}^2,
\en(12)
$$
where $D$ is a dimension of the system. This formula expresses $\sigma_{e}$
only through two first moments of random variable $\sigma.$
The further theoretical results in this problem for random inhomogeneous
systems are related with the effective medium approximation (EMA) \ci{9}
and the random resistor networks (RRN) models of these
systems \ci{2}, since they allow to fulfil an averaging procedure explicitly.
The RRN models approximate the continuous inhomogeneous systems (IS) and
can be obtained as a result of some their discretization (though, in general,
this procedure is not exact and unambiguous).
It was shown for the RRN models
firstly by numerical simulations \ci{2} and later by series expansion on
moments of $\delta \sigma$ \ci{10,11}
that the EMA, dealing  ad hoc with weakly IS,
works well even for  moderately inhomogeneous RRN models.

It is widely believed that the EMA, due to its good description of RRN type
models, is also a good approximation for continuous random IS.
As is known the EMA for different isotropic
RRN models (for example, on isotropic regular lattices) gives different
expressions for $\sigma_e$ \ci{2}. Thus $\sigma_e$ in the EMA is not an
universal function even for isotropic  RRN models.
At the same time, not every continuous random IS
can be described by the discrete RRN models, since their reduction to the
RRN type models can change some properties.
Moreover, the continuous random IS can have more complicate local
structures (see below). Then $\sigma_{e}$ of such continuous systems
can differ in higher order terms in $\delta \sigma$ (starting with the 3-rd
order, because the two first terms are universal and are determined by the
formula (12)) from the EMA. For example, it is not obvious that any system,
containing random "resistors" and "conductors" (i.e. the elements inversed
to each other) simultaneously, must have the same behaviour as a "pure" RRN
model. For a description of continuous IS it is very desirable to have any
method, basing on the first principles and general properties, which is not
connected with the RRN type models and/or with the
EM approximation. Fortunately, for two-dimensional systems,
due to the existence of the exact duality relation \ci{12,13},
one can attempt, basing on this relation, to find other possible
approximations different from the EMA and applicable for systems with other
inhomogeneity structures.
In this paper we will show for the binary systems ($N=2$)
how starting with
a duality relation (DR) and using some additional assumptions one can find
two explicit approximate expressions for $\sigma_{e},$ differing from the EMA.
We will also construct  two corresponding models, having their effective conductivities
just of these two forms, and discuss their properties and possible applications.
A case of arbitrary number of phases $N$ will be presented in the subsequent
paper.
\bs

\cl{\bf 2. Reciprocity and duality relations and  symmetries }

\bs

We begin with a consideration of main general properties of $\sigma_e,$
because in the existing literature there are some misunderstandings about
them. For systems with  $\delta$-correlated inhomogeneities $\sigma_e$
can be reduced to a function, depending on the pairs of partial parameters:
conductivities $\sigma_i$ and phase concentrations
$x_i \;(0 \le x_i \le 1, \; i=1,2, \; \sum_1^2 x_i =1)$
$$
\sigma_e =   \sigma_e (\sigma_1, x_1|\sigma_2, x_2).
\en(13)
$$
Note, that in general case the order of the arguments pairs is important,
since the phases can have different geometrical and/or statistical properties.
In two dimensions there is a reciprocity between divergenceless and curl free
vector fields ${\bf v}$ and $\tilde {\bf v}$
$$
v_i = \epsilon_{ik}\tilde v_k, \quad  \partial_i v_i =
\epsilon_{ik} \partial_i \tilde v_k = 0.
\en(14)
$$
 This relation is local and takes place in each point. It means that
the reciprocal vector field is the initial field turned  on $\pi/2.$
 For this reason for continuous inhomogeneous (and isotropic) media the
reciprocal vector field feels the same inhomogeneities of the initial system.
Inside any phase one can connect with the
reciprocal vector fields $\tilde {\bf j}$ and $\tilde {\bf E}$ a reciprocal
conducting system, such that  ${\bf j}_r \sim \tilde {\bf E}$ and ${\bf E}_r
\sim \tilde {\bf j},$ and for which the corresponding
Ohm law takes place
$$
{\bf j}_r = \tilde \sigma  {\bf E}_r, \quad
\tilde \sigma = \sigma_0^2 /\sigma,
\en(15)
$$
where $\sigma_0$ is some arbitrary constant with dimension of conductivity.
It means that $\tilde \sigma$ is inversed to $\sigma$ and the coordinate
axis of the reciprocal system are turned on $\pi/2.$
Note that the corresponding phase boundary conditions (2) are also fulfilled,
since the normal vectors transform into tangent ones and vice versa.
The  effective conductivities
of the initial medium and its reciprocal must satisfy  after averaging
the same functional relation as in (15), which can be called the reciprocity
relation (RR) and is usually written in the
following form
$$
\sigma_e(\sigma_1,x_1|\sigma_2,x_2)
\tilde \sigma_e(\sigma_0^2/\sigma_1,x_1|\sigma_0^2/\sigma_2,x_2)
= \sigma_0^2.
\en(16)
$$
In this general form it  have been
obtained firstly  by Keller \ci{12} and independently by Dykhne \ci{13}.
The general RR (16) takes place for all two-dimensional  binary systems:
for RRN models and for regular and irregular media, as well as for random ones.
But the general RR is not so useful, since it connects
the effective conductivities of different systems.
In order to obtain more information about $\sigma_e$ one needs to know how
these effective conductivities are related. The most interesting case is
when a reciprocal system has the same statistical or/and geometrical properties.
In this case the effective conductivity of
a reciprocal system must coincide with the effective conductivity of the initial one.
Such systems can be called the self-reciprocal ones
(we will use  a name self-dual for more symmetrical systems, see below).
One can see from the above description of the reciprocal medium that for
isotropic continuous random inhomogeneous medium
the reciprocal medium will have the same type of inhomogeneities and after
averaging its effective conductivity $\tilde \sigma_e$ must have the same functional form
as the effective conductivity of the initial medium, but with
inversed partial conductivities. Thus they have satisfy
the self-reciprocity relation (SRR)
$$
\sigma_e(\sigma_1,x_1|\sigma_2,x_2)
\sigma_e(\sigma_0^2/\sigma_1,x_1|\sigma_0^2/\sigma_2,x_2)
= \sigma_0^2.
\en(17)
$$
At the same time it is worth to note here that in case of discrete RRN type
models the reciprocal network can differ from the initial one and consequently a
functional form of the effective
conductivity of the reciprocal network will differ from that of initial network.
In particular, the effective conductivities of RRN models on isotropic regular
lattices (let us call them random resistor lattice (RRL) models),
triangle and honeycomb, turn out different and must satisfy,
due to a mutual reciprocity of these lattices, only the R-relation
$$
\sigma_e^{tr} (\sigma_1, x_1|\sigma_2, x_2)
\sigma_e^{hon} (\sigma_0^2/\sigma_1, x_1|\sigma_0^2/\sigma_2, x_2) =
\sigma_0^2,
\en(18)
$$
This is a payment for an ambiguity of the discretization procedure of the
continuous medium.

The self-reciprocal systems include the continuous
media and RRN model on self-reciprocal networks. For self-reciprocal systems
the SRR can be rewritten also in the form
$$
\sigma_e (\sigma_1, x_1|\sigma_2, x_2) =
\sigma_0^2/
\sigma_e (\sigma_0^2/\sigma_1, x_1|\sigma_0^2/\sigma_2, x_2).
\en(19)
$$
This form  has a simple physical sense, connected with two possibilities of
a writing the Ohm law in terms of conductivities $\sigma$ (as (15)) and in
terms of resistivities $\rho = 1/\sigma$:
an effective conductivity of a self-reciprocal system as a function of  partial conductivities
must coincide with an inversed resistivity as a function of corresponding
partial resistivities (i.e. inversed partial conductivities).
For this reason the relation (19) can be also called the inversion relation (IR).

Though now the SRR (17) restricts a possible functional form of $\sigma_e$,
in a general case it does not yet permit to find the exact values
for $\sigma_e$ even at equal phase concentrations
$x_1 =x_2 =1/2$ and $\sigma_0^2 = \sigma_1\sigma_2$ (the latter choice turns
the inversion $\sigma \to \sigma_0^2/\sigma$ into permutation
$\sigma_1 \lra \sigma_2$).
To be able to find exact value
of $\sigma_e$ one needs additional symmetry property of systems
connected with the geometrical and/or statistical properties of their
partial phases.
To see various possibilities, appearing for inhomogeneous media, let us
consider a simple, but very instructive, case of regular
inhomogeneous system consisting from one (white) phase with regular round
inclusions of the second (black) phase, the Rayleigh model.
This system is self-reciprocal in the above sense
and its effective conductivity satisfies the SRR (and IR) (17)
(\ci{14}). However, the geometrical properties of these phases are different.
The four pictures of this model at the interchanged phases and at the adjoint
concentrations $x$ and $1-x$ of black phase are represented on Fig.1.
\begin{figure}[t]
\begin{picture}(250,240)
\put(50,100){%
\begin{picture}(80,80)
\put(10,10){\line(0,1){60}}
\put(10,10){\line(1,0){60}}
\put(70,10){\line(0,1){60}}
\put(10,70){\line(1,0){60}}
\multiput(20,20)(13,0){4}{\circle*{6}}
\multiput(20,33)(13,0){4}{\circle*{6}}
\multiput(20,46)(13,0){4}{\circle*{6}}
\multiput(20,59)(13,0){4}{\circle*{6}}
\put(32,-8){({\small a})}
\end{picture}}

\put(50,0){%
\begin{picture}(80,80)
\put(10,10){\line(0,1){60}}
\put(10,10){\line(1,0){60}}
\put(70,10){\line(0,1){60}}
\put(10,70){\line(1,0){60}}
\put(32,-8){({\small c})}
\end{picture}}

\put(200,100){%
\begin{picture}(80,80)
\put(10,10){\line(0,1){60}}
\put(10,10){\line(1,0){60}}
\put(70,10){\line(0,1){60}}
\put(10,70){\line(1,0){60}}
\multiput(20,20)(13,0){4}{\circle*{14}}
\multiput(20,33)(13,0){4}{\circle*{14}}
\multiput(20,46)(13,0){4}{\circle*{14}}
\multiput(20,59)(13,0){4}{\circle*{14}}
\put(32,-8){({\small b})}
\end{picture}}

\put(200,0){%
\begin{picture}(80,80)
\put(70,10){\line(0,1){60}}
\put(10,70){\line(1,0){60}}
\put(10,10){\line(0,1){60}}
\put(10,10){\line(1,0){60}}
\put(32,-8){({\small d})}
\end{picture}}
\end{picture}
\caption{\small (a) A part of the model with  regularly situated
round inclusions at the concentration $x$, (b) a part of the same model
at adjoint concentration $1-x$, (c) the same model at concentration $x$
but with the interchanged phases,
(d) the same model with interchanged phases at adjoint concentration $1-x$.}
\end{figure}
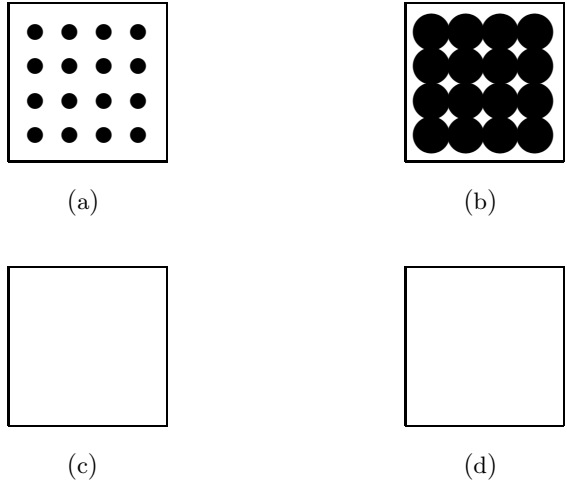
As follows from Fig.1,  all these examples of the model are different and,
consequently, the corresponding $\sigma_e$ are also different.
Thus, though the SRR takes place for this model, it does not give any
equation for $\sigma_e$.
Such equation can be obtained only when $\sigma_e$ is a symmetrical
under simultaneous interchange of both parameters of phases
$$
\sigma_e (\sigma_1, x_1|\sigma_2, x_2) =
\sigma_e (\sigma_2, x_2|\sigma_1, x_1).
\en(20)
$$
This property can take place as for random systems as for some regular
systems. The typical examples of such systems are
the random mixtures and RRN models with the probability distribution function
$P(\sigma)$ (7), and  a regular "chess-board"
type system. One can see from Fig.1 that the self-reciprocal
Rayleigh model does not  have this symmetry, but this symmetry appears as
an approximate symmetry, when a concentration of one phase is small (then
fig.1a (fig.1b) is similar to fig.1d (fig.1c)).

We will call symmetrical self-reciprocal systems as self-dual (SD) systems.
The corresponding SRR takes at $\sigma_0^2 = \sigma_1 \sigma_2$  the
following form
$$
\sigma_e(\sigma_1,x_1|\sigma_2,x_2)
\sigma_e(\sigma_1,x_2|\sigma_2,x_1)
= \sigma_1 \sigma_2.
\en(21)
$$
For SD systems one can obtain exact equation
for $\sigma_e.$ It takes place only
at equal phase concentrations $x_i=1/2$ and at $\sigma_0^2 = \sigma_1 \sigma_2.$
and gives the well-known exact Keller-Dykhne value \ci{12,13}
$$
\sigma_e (\sigma_1, 1/2|\sigma_2, 1/2) =
\sqrt{\sigma_1 \sigma_2}.
\en(22)
$$
This remarkable formula is very simple and universal in a sense that it
takes place for all two-dimensional inhomogeneous self-dual systems
as for regular as for random.

For binary systems one can write $\sigma_e$ in a more short form as a
function  of its independent parameters only (here $x$ is a concentration
of the first phase)
$$
\sigma_e = \sigma_e(x,\sigma_1,\sigma_2).
\en(23)
$$
Then, for isotropic SD systems with unequal phase concentrations (21) can
be written in the form, which connects the effective conductivities
at adjoint concentrations $x$ and $1-x$ or in terms of a new variable
$\epsilon = x-1/2$ ($ -1/2 \le \epsilon \le 1/2$) at $\epsilon$ and
$-\epsilon$
$$
\sigma_{e}(x, \sigma_1, \sigma_2 )
\sigma_{e} (1-x, \sigma_1, \sigma_2) = \sigma_1 \sigma_2
\en(24)
$$
Eq.(24) means that a product of the effective conductivities at adjoint
concentrations is an invariant. Due to this relation one can consider
$\sigma_{e}$ only in the regions $x \ge 1/2$ ($\epsilon \ge 0$) or
$x \le 1/2$ ($\epsilon \le 0$).
Just the relations (21) and (24) are usually named as the duality relation
(DR). Below we will consider mainly systems,
satisfying the DR (21) or (24) as the exact functional relation.
It follows also from (24) that (for $\sigma_1 \ge \sigma_2$) than faster
is $\sigma_e$ increasing at $x \ge 1/2,$ than slower it increase at $x \le
1/2.$

The binary case  assumes further simplifications.
It is convenient to use instead of $\sigma_i, i=1,2$
another combinations of $\sigma_i$: $\sigma_{\pm} = (\sigma_1 \pm \sigma_2)/2.$
Then, the effective conductivity can be represented in the form, symmetrical
relatively to both phases,
$$
\sigma_{e}(\epsilon, \sigma_+, \sigma_- ) =
\sigma_{+} f(\epsilon, \sigma_-/ \sigma_+) =
\sigma_{+} f(\epsilon, z),
\en(25)
$$
where a new dimensionless variable $z=\sigma_{-}/ \sigma_+,\; (-1 \le z \le 1),$
 characterizing an inhomogeneity of the system, is introduced. This choice
of variables is convenient for a study of $\sigma_e$ in the limit, when one
partial conductivity becomes small or  goes to 0.

The effective conductivity
$\sigma_{e}(\epsilon, \sigma_+, \sigma_- )$ and a function $f(\epsilon, z)$
must have the next boundary values
$$
\sigma_{e}(1/2, \sigma_+, \sigma_- ) = \sigma_1, \quad
\sigma_{e}(-1/2, \sigma_+, \sigma_- ) = \sigma_2,
$$
$$
f(1/2, z) = 1+z, \quad f(-1/2, z) = 1-z, \quad f(\epsilon, 0) =1.
\en(25')
$$
The duality relation takes in these variables the form
$$
f(\epsilon, z) f(-\epsilon, z) = 1-z^2,
\en(26)
$$
from which one obtains  at critical concentration $\epsilon = 0$ the exact
value
$$
f(0, z) = \sqrt{1-z^2}.
\en(26')
$$
The self-reciprocity relation in these terms has  the form
$$
f(\epsilon, z) f(\epsilon, -z) = 1-z^2,
\en(27)
$$
and cannot give a nontrivial exact value, since $f(\epsilon, 0) = 1.$
The symmetry (20) in the new variables means that
$$
f(\epsilon, z) = f(-\epsilon, -z),\quad f(-\epsilon, z) = f(\epsilon, -z).
\en(28)
$$
It follows from (28) that the even ($f_s$) and odd ($f_a$) parts of
$f(\epsilon,z)$ relatively
to $\epsilon$  coincide with the even ($f^s$) and odd ($f^a$) parts of
$f(\epsilon,z)$ relatively to $z.$
The simplest way to satisfy these properties corresponds to the
following functional form of $f(\epsilon,z)$
$$
f(\epsilon,z) = f(\epsilon z, \epsilon^2, z^2)
\en(29)
$$
One can conclude from (29) that in this case an expansion of $f(\epsilon,z)$
near the point $\epsilon = z = 0$ does not contain terms linear in
$\epsilon$ and $z$ separately.
The odd part $f_a$ can be represented in the form
$$
f_a(\epsilon,z) = 2\epsilon z \Phi(\epsilon, z)
\en(30)
$$
where $\Phi$ is an even function of $\epsilon$ and $z$ (the coefficient $2$
in front of $\epsilon z$ is chosen for further convenience).

The DR (24) in general case is not enough for a complete
determination of $f$, but it poses some constraint on the possible form of
$f.$ There are various ways of its choice. The most known example, using
the partition of $f$ into symmetric and antisymmetric parts $f = f_s + f_a,$
connects $f_s$ and $f_a$
$$
f_s^2 - f_a^2 = 1-z^2.
\en(31)
$$
 Since  this form of the DR is associated with "a parity" partition
of function $f$ one can call it  "a parity representation" of the DR.
It means that $f_a$ and $f_s$ considered at fixed $z$ as the functions
of $\epsilon$ satisfy to hyperbolic relation with a constant depending on
$z.$ The relation (31) allows to express $f(\epsilon,z)$ through its even
or odd parts
$$
f(\epsilon,z) =
f_a + \sqrt{f_a^2 + 1-z^2} =
f_s \pm \sqrt{f_s^2 - 1 + z^2}.
\en(32)
$$
For this reason it is enough to know only one of these two parts.
Usually one prefers to choose an antisymmetric part as more simple.
Using the first
approximation in $z$ for $f_a$ in the weakly inhomogeneous case,
which follows from the mean conductivity
$$
\langle \sigma \rangle = \sigma_+ (1+2 \epsilon z), \quad
f_a(\epsilon,z) = 2 \epsilon z.
\en(33)
$$
and  substituting (33) into (32), one obtains the effective conductivity
$$
\sigma_{e}(\epsilon,z) = \sigma_+ \left[2 \epsilon z +
\sqrt{(2 \epsilon z)^2 + 1 - z^2}\right],
\en(34)
$$
which coincides with $\sigma_{e}$ of self-dual square RRL model in the EMA \ci{2}.
We will refer the expression (34) as the EMA expression for effective
conductivity of self-dual systems. For self-reciprocal systems there are also
relations analogous to (31,32) with a change $f_s, f_a \to f^s, f^a,$ but
in this case the relations (29,30)do not take place. For this reason the EMA
cannot be a good approximation for SR systems.

For a checking the symmetry properties and a comparison
of different expressions for $\sigma_e$ it is useful to consider also
$\sigma_e$  of RRL models on other isotropic lattices, triangle and honeycomb.
The EMA expression for $\sigma_e$ of the binary lattice
RRN models is \cite{2}
$$
\sigma_e = \fr{1}{q-2}\left\{[]
+ \sqrt{[]^2 + 2(q-2)\sigma_1 \sigma_2}\right\},
$$
$$
[] = \left[\left(\fr{q}{2}x-1\right)\sigma_1 +
\left(\fr{q}{2}(1-x)-1\right)\sigma_2\right]
\en(35)
$$
where $q$ is a number of bonds coming in (or outgoing from) one lattice site.
In terms of $z$ and $\epsilon$ it has the following form
$$
\sigma_e = \fr{\sigma_+}{q-2}\left\{q\left(\fr{q-4}{2q} + \epsilon z\right)+
[]^{1/2}\right\},
$$
$$
[] = [(q/2)^2 + q(q-4)\epsilon z + (q^2 \epsilon^2-2(q-2)) z^2]
\en(36)
$$
which has the necessary symmetry property (29).
One can check that the corresponding $\sigma_e^{(tr)}$ (q=6) and
$\sigma_e^{(hon)}$ (q=3) satisfy also the RR
$$
\sigma_e^{tr}(\epsilon, z) \sigma_e^{hon}(-\epsilon, z)= \sigma_1 \sigma_2.
$$
As it was remarked earlier, some properties of RRN models differ from
the properties of continuous media.
For this reason, one can hope that the other possible representations for $f$
compatible with the DR can produce other approximate expressions
for $\sigma_e,$ which will turn out more suitable for models with other
inhomogeneity structures.
Another reason to analyse the duality relation is connected with
the known fact, that usually the systems with such symmetry have some
additional, hidden, properties, permitting to obtain more information
about function under consideration.
Moreover, in some simple cases they can help to solve problem exactly
(see, for example, \ci{15}).
Bearing this in mind, we will investigate various consequences of
the duality relation in more details.

\bs

\centerline{\bf 3. Duality, symmetries and perturbation expansions}

\bs
We start with the series expansion of $f(\epsilon, z)$ on $z.$
For every fixed
$z \; (0 \le z < 1)$ a function $f$ must be a monotonous function of
$\epsilon.$ Since a homogeneous limit $z \to 0$ is a regular point of $f$
it will be very useful to consider a series expansion of $f$ in $z$
$$
f(\epsilon, z) = \sum_0^{\infty} f_k(\epsilon) z^k/k!,
\en(37)
$$
where due to boundary conditions (25')
$$
f_0 = 1, \quad f_1(\epsilon) = 2 \epsilon.
\en(38)
$$
It is worth to note here that: 1) similar expansion was already considered
up to third order (see, for example, \ci{16}) and 2) the expansion (38)
is connected with the weak disorder expansion for RRN models in moments
of fluctuations $\delta \sigma$ \ci{11}. We will show that our expansion
is much simpler, since it treats the variables $z$ and $\epsilon$
independently.  From this point of view  the expansion on moments of
$\delta \sigma$ is an expansion
on rather complicated functions of $z$ and $\epsilon.$

Substituting the expansion (37) into (26) one obtains
the following results:

(1) in the second order in $z$ it reproduces the universal formula (12),
thus the latter can be considered as a consequence of the duality relation;

(2) in general case, there are the recurrent relations between $f_{2k}$ and
$f_{2k-1},$ corresponding to the connection (31) and expressing the even
terms through the odd ones;

(3) $f_{2k+1}(\epsilon)$ are odd polynomials of $\epsilon$ of degree
$2k+1$ and
$f_{2k}(\epsilon)$ are even polynomials of $\epsilon$ of degree $2k$
in agreement with (29).

Taking into account boundary conditions (25'), the exact value
(26') and the symmetry property (29), one can show that the coefficients
$f_{k}$ must have the next form
$$
f_{2k+1}(\epsilon) = \epsilon (1-4\epsilon^2) g_{2k-2}(\epsilon),
\quad k \ge 1,
$$
$$
f_{2k}(\epsilon) = (1-4\epsilon^2) h_{2k-2}(\epsilon),
\quad k \ge 1,
\en(39)
$$
where $g_{2k-2}(\epsilon)$ and $h_{2k-2}(\epsilon)$ are some even polynomials
of the corresponding degree, and free terms of $h_{2k-2}(\epsilon)$
coincide with the coefficients in the expansion of the exact value (26')
$$
\sqrt{1-z^2} = 1-z^2/2 -z^4/8 -z^6/16 -z^8/128 -z^{10}/256 + ...
\en(40)
$$
The parity representation gives the relations, expressing  $f_{2k}$ with even
indices through $f_{k}$ with lower indices. For example,
$$
f_4 = 4f_1 f_3 - 3f_2^2 = (1-4\epsilon^2)[(8g_0 + 12)\epsilon^2 - 3],
\en(41)
$$
$$
f_6 = 6f_1 f_5 - 15 f_2 f_4 + 10 f_3^2.
\en(42)
$$

It follows from (39) that $f_3$ is completely determined up to overall
factor number $g_0.$ Since $f_4$ is determined through lower
$f_k \quad (k=1,2,3)$
it is also determined by the coefficient $g_0.$
In the EM approximation the series (21) is defined completely and takes very
simple form, since in
this case  all $g_{2k-2} = 0 \; (k>1)$ and
$f_{2k}(\epsilon) \sim (1-4\epsilon^2)^k$.

It is useful also to write out the series expansions for other 2D isotropic
RRL models in the EM approximation.
The expansion of (36) in $z$ up to the 6-th order terms gives
$$
\sigma_e = \sigma_+ \left\{1 + 2\epsilon z - \fr{2}{q}(1- 4\epsilon^2) z^2
+\fr{4\epsilon (1-4\epsilon^2)(8-6q+q^2)}{q^2(q-2)} z^3 \right.
$$
$$
- \fr{4(1-4\epsilon^2)(q-2+2\epsilon^2(20-10q+q^2))}{q^3} z^4 +
$$
$$
+ \fr{8\epsilon
(1-4\epsilon^2)}{q^4(q-2)}(8-6q+q^2)(3(q-2)+2\epsilon^2(28-14q+q^2)) z^5 -
$$
$$
- \fr{16(1-4\epsilon^2)}{q^5}\left((q-2)^2 + 2\epsilon^2
(-112 + 112q-34q^2+3q^3) +\right.
$$
$$
\left.\left.+ 2\epsilon^4(672-672q+224q^2-28q^3+q^4)\right) z^6 \right\}
\en(43)
$$
It has the necessary structure following from the general properties. Note
that the two first terms coincide with the universal (12) only for self-dual
square RRL.
One can see also from (43) that the higher odd terms (starting from the 3-rd order term)
depend on the lattice structure and do not
vanish except the self-dual case of the square lattice
(where this takes place due to the factor $(8-6q+q^2)$).

In order to find higher order terms for square RRL model one may
compare the expansion (37) with known results in series expansion in powers
of conductivity fluctuations $\delta_k = \langle (\delta \sigma)^k \rangle$
\ci{11}. Its coefficients are found up to 6-th
order and it was shown that for 2D square RRL model, due to the
duality relation, they begin to differ from the corresponding EMA coefficients
only in the 5-th order
$$
\sigma_e^{sq} = \langle \sigma \rangle \left(
1- \fr{\delta_2}{2} + \fr{\delta_3}{4} - \fr{\delta_4}{8} +
\fr{\delta_5+J_1\delta_2\delta_3}{16}\right.
$$
$$
\left.-\fr{[\delta_6+(3J_1-1)\delta_2\delta_4+(2J_1-1)\delta_3^2-
(3J_1-2)\delta_2^3]}{32}\right),
\en(44)
$$
where $J_1 =\simeq 1.092958179$
is some lattice integral of the square lattice
Green function.
The EM approximation corresponds to $J_1=1.$

For a comparison of the general expansion (44) with (37) one needs the
explicit expressions for the higher moments of $\delta \sigma$.
After some algebra one can show that the moments
of binary systems with the distribution function (7)
have in terms of variables $\sigma_-$ and $x$ (or $z$ and $\epsilon$)
the following form
$$
\langle (\Delta \sigma)^k \rangle = x(1-x) P_{k-2}(x) (\sigma_-)^k =
(\sigma_+)^k (1-4\epsilon^2)/4 P_{k-2}(\epsilon) z^k,
$$
$$
P_{k-2}(x) = (1-x)^{k-1} + (-1)^k x^{k-1},
$$
$$
P_{k-2}(\epsilon) = (1/2 - \epsilon)^{k-1} + (-1)^k (\epsilon +1/2)^{k-1}
\en(45)
$$
One can see from (45) that $z$-dependence of the moments is factorised.
Moreover, it follows from the explicit expression for the polynomial
$P_{k-2}$ that for
odd $k=2m+1$, due to the symmetry of this polynomial relative interchange
$x \leftrightarrow 1-x,$ it has additional general factor $(1-2x)= -2\epsilon$
in accordance with (39). Thus, we see that the series expansion for square
RRL model in $\delta_k$
from \ci{11} has in terms of variables $z$ and $\epsilon$ the necessary
structure, following from symmetry and duality. The obvious advantage
of series expansion (38) is that it treats the variables $z$ and
$\epsilon$ separately.
The explicit forms of the polynomials $P_{k-2}$ at $k \le 6$ are
$$
\langle \Delta^3 \rangle = 4 \epsilon (1-4\epsilon^2) \sigma_{-}^3, \quad
\langle \Delta^4 \rangle = 4 (1-4\epsilon^2)(1-3x +3x^2) \sigma_{-}^4,
$$
$$
\langle \Delta^5 \rangle = 4 \epsilon (1 - 4 \epsilon^2)(1 - 2x + 2x^2)
\sigma_{-}^5,
$$
$$
\langle \Delta^6 \rangle = 2^4 (1- 4\epsilon^2) (1- 5x + 10 x^2 - 10 x^3 +5x^4)
\sigma_{-}^6.
\en(46)
$$
Using the equality (33) for $\langle \sigma \rangle$ and expanding the terms
of the series (44) in $z,$ one can obtain the series (37)
for the square RRL model up to 6-th order. In particular, in
the EMA one obtains the corresponding series (43). Since the series for square
RRL model begin to differ from the corresponding EMA in 5-th order it is enough to
find only 5-th and 6-th order terms. They are
$$
g_2 = -30j(1-4\epsilon^2), \quad h_4 = -45(1-4\epsilon^2)
[1-4(1-2j)\epsilon^2],
\en(47)
$$
where $j=J_1 - 1 = 0.092958179 \ll 1.$
One can check that these formulas satisfy relations (41,42).

Thus, we have obtained the series in $z$ for $\sigma_e$  of self-dual square RRL
model up to 6-th order. Since it coincide with the EMA up to 4-th order,
one can say that they belong to the same class of functions with $g_0 =0.$
As follows from (43) the effective conductivity for isotropic RRL models
in the EM approximation depends on the lattice structure and can have $g_0
\ne 0.$ Then the next question
appears: is the effective conductivity of the continuous self-dual systems
universal and always belongs to the EMA class or there
are SD systems with other functional forms of $\sigma_e$? The answer on this
question will be obtained in the subsequent sections.

\bs

\cl{\bf 4. Exponential representation and new approximate expressions}

\bs
Thus we have seen that the arbitrariness of $f$ of self-dual systems is
strongly reduced by boundary conditions and the exact value (26'), and
that the third and fourth order terms are determined only up to one constant
$g_0,$ which equals 0 for the square RRL model and in the EM approximation.
The question about the universality reduces in this particular case to
the next one: is the equality $g_0 = 0$ true for all self-dual random inhomogeneous
systems or it depends on the structure of random inhomogeneities?
If the value of $g_0$ depends on random structure, then  the second
question appears: can any other
additional information about the function $f$ of these systems determine
the constant $g_0$ or even the whole function, as it was in case of the EMA?
Since the EMA is connected with the parity representation (31) and a mean
conductivity approximation (33) for odd part, one needs to know what kind of
other representations for $f$ can satisfy the duality relation (26).
In order to answer on this question
it is convenient in the case $z \ne 1$ to pass from $f$ to
$\tilde f = f/\sqrt{1-z^2}.$ Then
$$
\tilde f(\epsilon, z) \tilde f(-\epsilon, z) = 1 =
\tilde f(\epsilon, z) \tilde f(\epsilon, -z).
\en(48)
$$
The DR (48) assumes other possible functional forms of $\tilde f(\epsilon,z).$
Taking into account (29), one can write out
the  exponential representation for $\tilde f$
$$
\tilde f(\epsilon,z) = \exp(\epsilon z \phi(\epsilon,z)),
\en(49)
$$
where $\phi(\epsilon, z)$ is some even function of its arguments.
It is easy to see that it automatically satisfies eq.(48).
Now let us  find out is it possible to define $g_0$ from some additional
information about function $\phi.$  For this we consider two simple ansatzes
for $\phi(\epsilon, z)$, analogously to the parity representation (31)
and the EM approximation.

1) In the first case
we will assume that $\phi(\epsilon, z)$
depends only on $z.$ This corresponds to the exponential dependence on
concentration. Such dependence takes place often in disordered systems \ci{17}.
Expanding the corresponding functions $\tilde f$ in series one can check
after some algebra that now  all polynomial coefficients can be determined
unambiguously!
The series for $f_a$ up to 5-th order has the form
$$
f_a = \sum_0^2 f_{2k+1} \frac{z^{2k+1}}{(2k+1)!}, \quad
f_{2k+1} = \epsilon(1-4\epsilon^2)g_{2k-2}, \quad k=1,2,
$$
$$
g_0 = -1, \quad g_2 = -(11+4\epsilon^2).
\en(50)
$$
Another way to see this is to apply the boundary conditions (25') directly
to the function (49). Then one obtains the whole function
$$
\phi(z) = 1/z \ln \fr{1+z}{1-z}, \quad
\tilde f (\epsilon,z) =  \left(\fr{1+z}{1-z}\right)^{\epsilon}.
\en(51)
$$
The series expansion of (51) reproduces exactly the corresponding
expansion from (50). Note that now $g_0 \ne 0.$ The even part $f_s$ can be
found from (41,42).

2) In the second case
we will assume that $\phi(\epsilon, z)$ depends only on the combination
$\epsilon z.$ This corresponds to the dependence of $\sigma_e$ only on
mean conductivity and/or mean resistivity, since
$\langle \sigma^{\pm 1} \rangle \sim (1\pm 2\epsilon z).$
Expanding the corresponding functions $\tilde f$ in series one can see
that again  all polynomial coefficients can be determined completely.
The coefficients for $f_a$ up to 5-th order are
$$
g_0 = -3, \quad g_2 = -15(1+ 12\epsilon^2)
\en(52)
$$
In this case one can also apply the boundary conditions (25') directly to the
function (49). Then one obtains
$$
\phi(\epsilon, z) = \fr{1}{2\epsilon z} \ln \fr{1 + 2\epsilon z}{1-2\epsilon z},
\quad \tilde f(\epsilon,z) =
\left(\fr{1+ 2\epsilon z}{1-2\epsilon z}\right)^{1/2}.
\en(53)
$$
The series expansion of (53) also coincides exactly with the corresponding
expansion represented in (52).

It is worth to note that an analogous exponential representation can be
obtained for self-reciprocal, but nonsymmetric systems.
In this case only the second equality in (48) is fulfilled and the
exponential representation has the form
$$
\tilde f(\epsilon,z) = \exp(z \phi(\epsilon,z)),
\en(54)
$$
where $\phi(\epsilon,z)$ is an odd function of $z.$ The boundary conditions
(25') cannot now determine it completely. They only give
$$
\phi(\pm \frac{1}{2},z) = \pm \frac{1}{2z} \ln \left(\frac{1+z}{1-z}\right).
$$
The simplest way to satisfy these boundary conditions is to assume that the
interpolating function at arbitrary $\epsilon$ has factorized dependencies
on $\epsilon$ and $z$ (it is an analog of the  assumption 1) in the SD case)
$$
\phi(\epsilon,z) = \frac{\alpha(\epsilon)}{z} \ln \left(\frac{1+z}{1-z}\right),
\en(55)
$$
here  $\alpha(\epsilon)$ is an arbitrary  function except its boundary
values
$$
\alpha(\pm 1/2) = \pm 1/2.
$$
For self-dual systems $\alpha(\epsilon) = \epsilon,$ and just this equality
ensures the exact Keller-Dykhne value (22,26') at $\epsilon =0.$
Formally, the SR relation admits the self-dual solution (52) in nonsymmetric
case too, but, as we will show below in section 7, (52) works  well only for
$\epsilon,$ belonging to small regions near the boundary values
$\epsilon = \pm 1/2$ (or at small concentrations of one of the phases
$x \ll 1, 1-x \ll 1$).
Its applicability at intermediate values of $\epsilon$ depends on how close
is a non self-dual system to the self-dual one.

Thus  we have shown that under some simple (but physically acceptable)
additional assumptions about a function $f(\epsilon, z)$ one can obtain
the new approximate (a mean field like) expressions for $f.$
Their physical meaning and properties will be discussed in two subsequent
sections  where we will present two simple models having
the effective conductivity just of two  forms found above.
Here we only want to note that it is very important
for further progress in this direction to find any other explicit solutions.
In general case, assuming an expansion of $\phi$ in double series in $z$ and
$\epsilon,$
$$
\phi(\epsilon,z) = \sum_{k,l = 0}^{\infty} \phi_{kl}
\frac{z^{2k}}{k!} \frac{\epsilon^{2l}}{l!}
$$
one can show that now again the
boundary conditions (25') cannot define all coefficients completely.
For example, $f_3$ and $f_4$ contain one free parameter $\phi_{10}:$
$g_0 = 6(\phi_{10} -1).$

\bs
\cl{\bf 5. Finite maximal scale averaging approximation}
\bs

In this section we show that the solution (52) can describe $\sigma_e$
of inhomogeneous systems with compact inclusions of one of the phases.
Firstly, we introduce a new composite method of a construction of random
inhomogeneous systems. In framework of this approach one divides an averaging
procedure on two steps, corresponding to the averaging on two different scales.
Then, using this approximate averaging,
which we will call as a finite maximal scale averaging approximation (FMSA),
we will obtain an additional equation for effective conductivity of
inhomogeneous systems with compact inclusions.
This approximation can be considered as some modification
of the mean field approximation and it is applicable to the random
inhomogeneous systems with a finite maximal scale of the inhomogeneities.
Below in this section we will use temporally, for clarity,
the concentration variable $x$ instead of $\epsilon.$
Let us consider planar two-phase randomly inhomogeneous system.
It is easy to see that, in general, its effective conductivity will depend
on a scale $l$ of a region over which an averaging is done. This takes place
due to the possible existence of different characteristic scales in the
inhomogeneous medium. In the most general case there will be a whole spectrum
of these characteristic scales. This spectrum can change from
discrete finite one till continuous infinite one and will define the
inhomogeneity structure of the system. For this reason this spectrum can
depend on concentration $x.$ It is obvious that the effective conductivity
of the system can depend on this spectrum as a whole as well as on
which area an averaging is fulfilled over.
Suppose, for the simplicity, that the randomly inhomogeneous structure of
our system has a scale spectrum with the maximal scale $l_m(x),$ which is
finite for all $x$ in the region $1 \ge x > 1/2$ (or $1-x$ in the region
$0 \le 1-x < 1/2$). Let us assume that we know an
exact formula for $\sigma_{e}(x,z)$ of this system, which is applicable from
scales $l > l_m.$
It means that this formula for $\sigma_{e}(x,z)$ takes place after the
averaging over regions with a mean size $l \gtrsim l_{m}$ and does not
change for all larger scales $l \gg l_m.$
Now we deduce an approximate functional equation for this $\sigma_{e}(x,z).$
To do this let us consider the auxiliary square lattice
with the squares of length $l_L \gg l_{m}$ and suppose that they have the
effective conductivities corresponding to different values of the
concentrations $x_1$ and $x_2$ with equal probabilities $p = 1/2$ (see Fig.2)

\begin{figure}
\begin{picture}(250,120)
\put(45,20){\line(0,1){50}}
\put(45,20){\line(1,0){50}}
\put(95,20){\line(0,1){50}}
\put(45,70){\line(1,0){50}}
\put(55,30){\circle*{15}}
\put(60,55){\circle*{10}}
\put(75,60){\circle*{20}}
\put(85,40){\circle*{10}}
\put(65,45){\circle*{5}}
\put(85,35){\circle*{5}}
\put(80,55){\circle*{5}}
\put(54,35){\circle*{5}}
\put(60,35){\circle*{5}}
\put(67,43){\circle*{5}}
\put(65,5){({\small a})}
\put(130,0){%
\begin{picture}(100,50)%
\multiput(75,20)(10,0){11}%
{\line(0,1){50}}
\multiput(75,20)(0,10){6}%
{\line(1,0){100}}
\multiput(77.5,21)(0,20){3}{1}
\multiput(77.5,31)(0,20){2}{2}
\multiput(87.5,31)(0,20){2}{1}
\multiput(87.5,21)(0,20){3}{2}
\multiput(97.5,21)(0,20){3}{1}
\multiput(97.5,31)(0,20){2}{2}
\multiput(107.5,31)(0,10){3}{2}
\multiput(107.5,21)(0,40){2}{1}
\multiput(117.5,51)(0,10){2}{2}
\multiput(117.5,21)(0,10){3}{1}
\multiput(127.5,21)(0,10){5}{2}
\multiput(137.5,21)(0,20){3}{1}
\multiput(137.5,31)(0,20){2}{2}
\multiput(147.5,21)(0,10){5}{1}
\multiput(157.5,31)(0,10){2}{2}
\multiput(157.5,21)(0,30){2}{1}
\multiput(167.5,31)(0,20){2}{2}
\multiput(167.5,21)(0,20){2}{1}
\put(167.5,61){2}
\put(157.5,61){2}
\put(120,5){({\small b})}
\end{picture}}
\end{picture}
\caption{\small (a) An elementary square of the auxiliary lattice
with compact inclusions, (b) the auxiliary lattice, the numbers 1,2 denotes
squares with the corresponding concentrations $x_1$ and $x_2$.}
\end{figure}
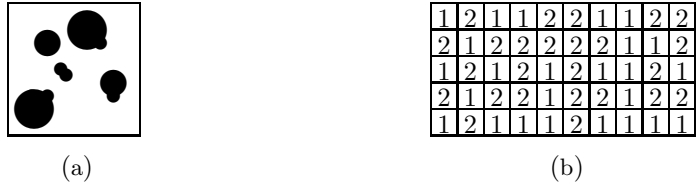

After the averaging over the scales $l \gtrsim l_L$ one must obtain on much
larger scales $l \gg l_{L}$ the same effective conductivity,
but corresponding to another concentration $x=(x_1 + x_2)/2.$
This is possible due to the similar random structure of different squares
and due to the conjectures that in this
model: (1) there are only two maximal characteristic scales
$l_m(x_i) \equiv l_i \quad (i=1,2)$ and the lattice square size
$l_L \gg l_i, l_m(x)$, (2) the averaging procedures over these scales do not
correlate (or weakly correlate) between themselves. Thus, for a compatibility,
all concentrations must be out of small region around critical concentration
$x_c,$ where $l_i$ or $l_m(x)$ can be very large. We will call further
this set of the assumptions the finite maximal scale averaging approximation
(FMSA approximation). It can be implemented for systems with compact
inhomogeneous inclusions with finite $l_m$.
From the other side the effective conductivity of the auxiliary plane
on scales $l \gg l_L$ must be determined  by the universal
Keller -- Dykhne formula (22). Thus we obtain the next functional equation
for the effective conductivity, connecting $\sigma_{e}(x,z)$ at different
concentrations,
$$
\sigma_{e}(x,z) = \sqrt{\sigma_{e}(x_1,z) \sigma_{e}(x_2,z)},\quad
x=(x_1+x_2)/2, \quad (x_i,x \ne x_c).
\en(56)
$$
It must be supplemented by the boundary conditions (25').
This equation can be considered as a generalization of the duality relation
(21), the latter being a particular case of (56) at $x_1 + x_2 = 1.$
It follows from (56) that, due to the exactness of the duality relation,
it really works for all concentrations $x,$ except, maybe, a small region
near $x=x_c$ and $z=1$ (this region corresponds to the singular region of
the percolation problem, see also below a discussion of the percolation limit).
One can easily check that $\sigma_{e}$ in low concentration limit, being
a linear function of $x,$ satisfies (56). Moreover, one can find an exact
solution of this equation. It has an exponential form
with a linear function of $x$ (or $\epsilon$)
$$
\sigma_{e}(x,z) = \sigma_{1} \exp(ax + b),
\en(57)
$$
where the constants $a,b$ can be determined from the boundary conditions
$$
a = - b, \quad \exp b = \sigma_2/\sigma_1.
\en(57')
$$
Substituting these coefficients into (57) one obtains
$$
\sigma_{e}(x,z) = \sigma_{1} \left(\sigma_2/\sigma_1\right)^{(1-x)}
= \sigma_+\sqrt{1-z^2}\left(\frac{1+z}{1-z}\right)^{\epsilon}.
\en(58)
$$
The solution (58) satisfies symmetry relation (20) and can be
represented in the exponential form
$$
\sigma_{e}(x,z) =
\sigma_+\sqrt{1-z^2}\exp\left(\epsilon \ln \fr{1+z}{1-z}\right),
\en(58')
$$
which  coincides exactly with the case 1) from the section 4 with
$\phi(\epsilon,z) = \fr{1}{z} \ln \fr{1+z}{1-z}$.
Its odd part has a form
$$
\tilde f_a(\epsilon,z) = \sinh (\epsilon z \phi(\epsilon,z)),
\en(59)
$$
and automatically satisfies (31).

It is interesting to note that the form of the solution (58) means that
in the considered approximation one has effectively an averaging  of the
$\log \sigma$ since it can be represented as
$$
\log \sigma_{e} = \langle \log \sigma \rangle =
x \log \sigma_1 + (1-x)\log \sigma_2.
\en(60)
$$
This coincides with a note made in \ci{13} for the case of equal concentrations
$x=1/2$ and with analogous results obtained later for disordered systems in
the theory of weak localization \ci{18}.
The solution (58) satisfies also the inequalities (11) due to the
H\"older inequalities
$$
\langle \sigma^{-1} \rangle^{-1} \le
\sigma_{e}(x,z) = \sigma_{1}^{x} \sigma_2^{(1-x)}
\le \langle \sigma \rangle.
\en(61)
$$

One can check that (58) reproduces
in the weakly inhomogeneous limit the universal Landau -- Lifshitz expression
(12). In the low concentration limit of the second phase it gives
$$
\sigma_{e}(x,z) = \sigma_1(1 + (1-x)\log \fr{1-z}{1+z} + ...),
\quad 1-x \ll 1,
\en(62)
$$
what coincides with the Maxwell-Garnett formula  in the weakly
inhomogeneous case \ci{1}.
Note that the expansion (62) contains the coefficient logarithmically
diverging in the limit $|z| \to 1\; (\sigma_2 \to 0).$
Such behaviour of the coefficients denotes
the existence of a singularity in this limit as it must be due to the
percolation singularity (see below a discussion of this percolation limit).

As follows from the arguments of the method used above, the FMSA
 can be applied also for approximate
description not only of random systems, but also for any (regular as well
as irregular) inhomogeneous models with compact inclusions of one phase
into another. However, the explicit approximate expression (58) obtained for
$\sigma_e$ will work well in a wide
range of parameters only for systems close to the self-dual ones (see the notes
in the sections 4 and 7).
For example, this expression describes $\sigma_e$ of the Rayleigh
model with strictly round inclusions not very well for intermediate
concentrations (it gives smaller values) as it shows the comparison
from section 7 (it can be expected also from Fig.1). But it will work much
better when the shape of these inclusions is deformed so that the
corresponding pictures of this model (Fig.1a and Fig.1d, Fig.1b and Fig.1c)
become similar and  symmetry property (20) is approximately fulfilled.

\bs
\cl{\bf 6. Random "parquet" model}
\bs

Now we will show that the existence of different functional forms for the
effective conductivity denotes a nonuniversal character of this value for
self-dual two-phase systems with different random structures.
We will construct  another model of two-dimensional isotropic randomly
inhomogeneous two-phase system, using the composition method introduced above,
and find a mean field like expression for its effective
conductivity $\sigma_{e}(x,z)$ in case of arbitrary phase concentrations $x.$
The derived formula satisfies again to all necessary symmetries, including
a  self-dual one, and coincides with the example case 2) from section 4,
realizing the second variant, when $\tilde f(x,z)$ depends
(in this approximation) only on one combination of variables $\epsilon z.$
This dependence
is described by the  function analytical at small values of this variable.

Let us consider the following planar model. There is a simple
square lattice with the squares consisting of a random layered (or striped)
mixture of two
isotropic conducting phases with constant conductivities $\sigma_i, i = 1,2$
and the corresponding concentrations $x$ and $1-x.$  A schematic picture
of such square is given in Fig.3.

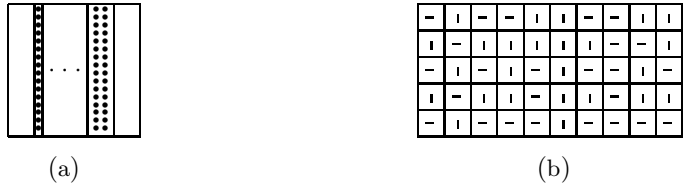
\begin{figure}
\begin{picture}(250,120)
\put(50,20){\line(1,0){50}}
\put(50,20){\line(0,1){50}}
\put(100,20){\line(0,1){50}}
\put(50,70){\line(1,0){50}}
\put(60,20){\line(0,1){50}}
\put(63,20){\line(0,1){50}}
\put(80,20){\line(0,1){50}}
\put(90,20){\line(0,1){50}}
\multiput(66.5,45)(5,0){3}{\circle*{1}}
\multiput(61.5,23)(0,3){16}{\circle*{2}}
\multiput(83.3,23)(0,3){16}{\circle*{2}}
\multiput(86.6,23)(0,3){16}{\circle*{2}}
\put(65,5){({\small a})}
\put(130,0){%
\begin{picture}(100,50)%
\multiput(75,20)(10,0){11}%
{\line(0,1){50}}
\multiput(75,20)(0,10){6}%
{\line(1,0){100}}
\put(120,5){({\small b})}
\multiput(78,25)(0,20){3}{\line(1,0){3}}
\multiput(80,33)(0,20){2}{\line(0,1){3}}
\multiput(88,35)(0,20){2}{\line(1,0){3}}
\multiput(90,23)(0,20){3}{\line(0,1){3}}
\multiput(98,25)(0,20){3}{\line(1,0){3}}
\multiput(100,33)(0,20){2}{\line(0,1){3}}
\multiput(110,33)(0,10){3}{\line(0,1){3}}
\multiput(108,25)(0,40){2}{\line(1,0){3}}
\multiput(120,53)(0,10){2}{\line(0,1){3}}
\multiput(118,25)(0,10){3}{\line(1,0){3}}
\multiput(130,23)(0,10){5}{\line(0,1){3}}
\multiput(138,25)(0,20){3}{\line(1,0){3}}
\multiput(140,33)(0,20){2}{\line(0,1){3}}
\multiput(148,25)(0,10){5}{\line(1,0){3}}
\multiput(160,33)(0,10){2}{\line(0,1){3}}
\multiput(158,25)(0,30){2}{\line(1,0){3}}
\multiput(170,33)(0,20){2}{\line(0,1){3}}
\multiput(168,25)(0,20){2}{\line(1,0){3}}
\put(170,63){\line(0,1){3}}
\put(160,63){\line(0,1){3}}
\end{picture}}
\end{picture}
\caption{\small (a) An elementary square of the model with a
vertical orientation, the dotted regions denote layers of the second phase;
(b)  a lattice of the model, the small lines on the
squares denote their orientations.}
\end{figure}

The layered structure
of the squares means that the squares have some preferred direction,
for example
along the layers. Let us suppose that the directions of different squares
are randomly oriented (parallely or perpendicularly) relative to the
external electric field, which is directed along $x$ axis.
In order for system to be isotropic the probabilities of the parallel and
perpendicular orientations of squares must be equal or (what is the same)
the concentrations of the squares with different orientations must be equal
$p_{||} = p_{\perp} = 1/2.$

Such lattice can model a random system consisting from mixed phase regions,
which can be roughly represented on the small macroscopic scales as randomly
distributed plots with the effective "parallel" and "serial" connections
of the layered two-phase mixture (Fig.3). The lines on the squares denote
their orientations.

This layered structure can appear,
for example, on the intermediate scales when a random medium is formed
as a result of the stirring of the two-phase mixture or as a result of
appearance of randomly distributed regions of "striped phases" in oxide
materials \ci{3}. The corresponding averaged parallel and
perpendicular conductivities of squares $\sigma_{||}(x)$ and
$\sigma_{\perp}(x)$ are defined by well known formulas
$$
\sigma_{||}(x) = \langle \sigma \rangle = x \sigma_1 + (1-x) \sigma_2 = \sigma_+ (1+ 2\epsilon z),
$$
$$
\sigma_{\perp}(x) = \langle \sigma^{-1} \rangle^{-1} = \left(\fr{x}{\sigma_1}+\fr{1-x}{\sigma_2}\right)^{-1}
= \sigma_+ \fr{1-z^2}{1-2\epsilon z}.
\en(63)
$$
Thus we have constructed the hierarchical random medium
(in this case a two-level one). On the first level it consists  from
some regions (squares of two different types) of the random mixture of the
two layered conducting phases with different conductivities
$\sigma_1$ and $\sigma_2$ and arbitrary concentration. On the second level
this medium is represented as a random parquet constructed from squares
of two such types with different conductivities $\sigma_{||}$ and
$\sigma_{\perp}$, depending nontrivially on concentration of the initial
conducting phases, and randomly distributed with the same probabilities
$p_i = 1/2$ (Fig.3).
This representation allows us to divide the averaging process approximately
into two steps (firstly averaging over each square and then averaging over
the lattice of squares) and implement on the second step the exact formula (22).
This can be considered as some modification of the FMSA approximation.
As a result one obtains for the effective conductivity of the introduced
random "parquet" model the following formula, which is applicable for
arbitrary concentration
$$
\sigma_{e}(\epsilon,z) = \sqrt{\langle \sigma \rangle \langle \sigma^{-1}\rangle^{-1}}
= \sigma_+ \sqrt{1-z^2} \tilde f(\epsilon, z),\quad
\tilde f(\epsilon, z) =
\left[\fr{1 + 2\epsilon z}{1 - 2\epsilon z}\right]^{1/2},
\en(64)
$$
This function has all necessary properties and satisfies SD equation (26),
inequalities (11) and the symmetry (20).
It coincides with the solution 2) from section 4 and has another possible
functional form, automatically satisfying the duality relation (48)
$$
\tilde f(\epsilon,z) = B(\epsilon,z)/B(-\epsilon,z)
\en(65)
$$
with a function
$B(\epsilon, z) = [1+2\epsilon z]^\fr{1}{2},$
which depends only on the combination $\epsilon z.$ Strictly speaking, this
form is not independent, since (65) can be also represented in the exponential
form
$$
\tilde f(\epsilon,z) = \exp(\phi(\epsilon,z))
$$
where $\phi(\epsilon,z)$ is an odd function of $\epsilon$
$$
\phi(\epsilon,z) = \ln B(\epsilon,z)/B(-\epsilon,z).
$$
It is interesting to compare this formula with the known asymptotic formulas.
Let us consider firstly its behaviour for small phase concentrations.

(a) In case of small concentration of the first phase $x \ll 1$ one gets
$$
\sigma_{e}(x,z) \simeq  \sigma_2 \left(1 + \fr{2xz}{1-z^2}\right).
\en(66)
$$
It follows from (66) that an addition of small part of the first  higher
conducting phase increases an effective conductivity of the system as it
should be.

(b) In the opposite case of small concentration of the second phase
$1 - x  \ll 1$ one obtains
$$
\sigma_{e}(x,z) \simeq  \sigma_1 \left(1 - \fr{2(1-x)z}{1-z^2}\right),
\en(67)
$$
i.e. an addition of the phase with smaller conductivity decreases
$\sigma_{eff}.$
It is worth to note that both these expressions
for arbitrary values of the conductivities $\sigma_1$ and $\sigma_2$
differ from the Maxwell-Garnett formula and coincide with it only in the
weakly inhomogeneous case $z \ll 1$.
It must be not surprising because  a form of the inclusions of the second
phase in this model has completely different, layered, structure.
In the low concentration expansion as well as
directly in formula (64) one can  see again that the divergencies appear
in the limit $z \to 1.$

(c) In case of almost equal phase concentrations
$x = 1/2 + \epsilon,$ $\epsilon \ll 1$ one obtains
$$
\tilde f(\epsilon,z) \simeq 1 + 2\epsilon z, \quad
\sigma_{eff}(\epsilon,z) \simeq \sigma_+ \sqrt{1-z^2} \left(1 +
2\epsilon z\right).
\en(68)
$$
The Keller -- Dykhne formula (22) is reproduced for equal concentrations.

The antisymmetric part of $\tilde f$ has the following form
$$
\tilde f_a(\epsilon,z) =
\fr{\tilde f^2(\epsilon,z)-1}{2\tilde f(\epsilon,z)} =
\fr{2\epsilon z}{\left[1 - 4\epsilon^2 z^2\right]^{1/2}}.
\en(69)
$$
It follows from formula (41) that for $\epsilon \ll 1$ and (or) $z \ll 1$
the odd part $f_a$ coincides in the first order with the corresponding
expression from the effective medium theory.
The corresponding function $\Phi$ is
$$
\Phi(\epsilon, z) =
\fr{\sqrt{1-z^2}}{\left[1 - 4\epsilon^2 z^2\right]^{1/2}}.
\en(70)
$$
In other words it is an analytic function of $\epsilon z$ near $\epsilon z=0$.
Basing on this formula one can conjecture that an exact expression for
$\tilde f_a(\epsilon,z)$ of this model will have a similar structure
in general case with $\sigma_i \ne 0$ or $1 - z \ne 1$.
One must note that at the same time the formula (64) does not satisfy the
equation (56) except of the trivial case $x_1 = x_2.$

\bs
\cl{\bf 7. Comparison of different approximations}
\bs

For the comparison of the different expressions of the effective conductivity
we have constructed 3D plots of $f(\epsilon,z)$ in
the EM approximation, in the FMSA approximation and for the
random "parquet" model in FMSA-like approximation (fig.4,5a).
For a clarity they are
represented at $z\le 0 (\sigma_1 \le \sigma_2).$

\begin{figure}[b]
\begin{tabular}{cc}
{\input epsf \epsfxsize=5.5cm \epsfbox{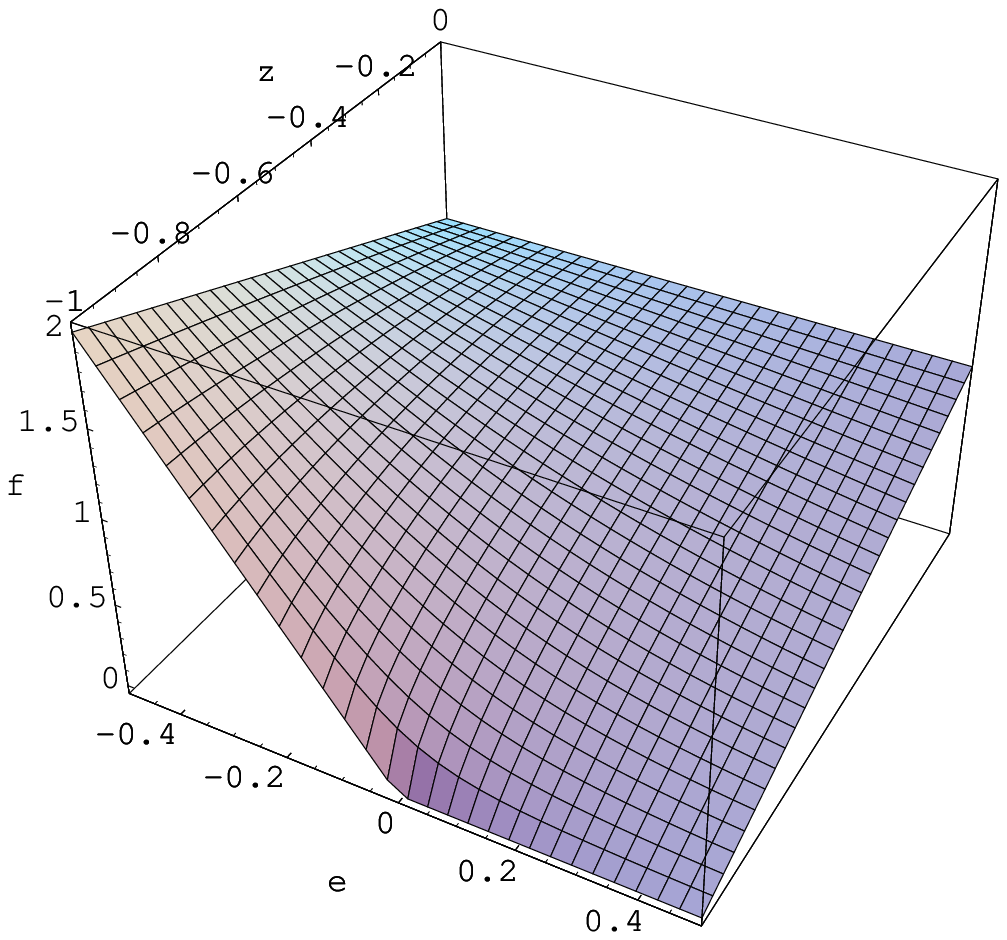}} &
{\input epsf \epsfxsize=5.5cm \epsfbox{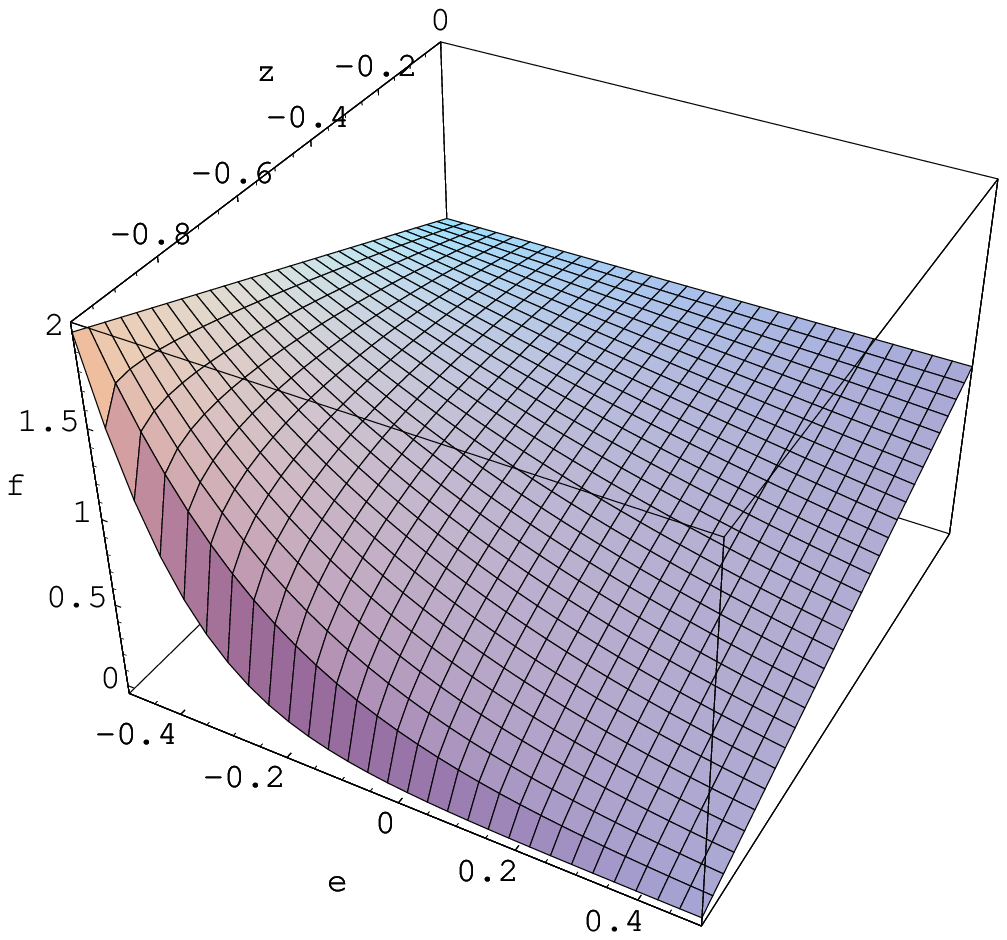}}\\
{} & {}\\
(a) & (b)\\
\end{tabular}
\vs{0.5cm}
{\small Figure 4: Plots of $f(\epsilon,z)$ in :
a) the EM approximation, b) the FMSA approximation.}
\end{figure}
\begin{figure}
\begin{tabular}{cc}
{\input epsf \epsfxsize=5.5cm \epsfbox{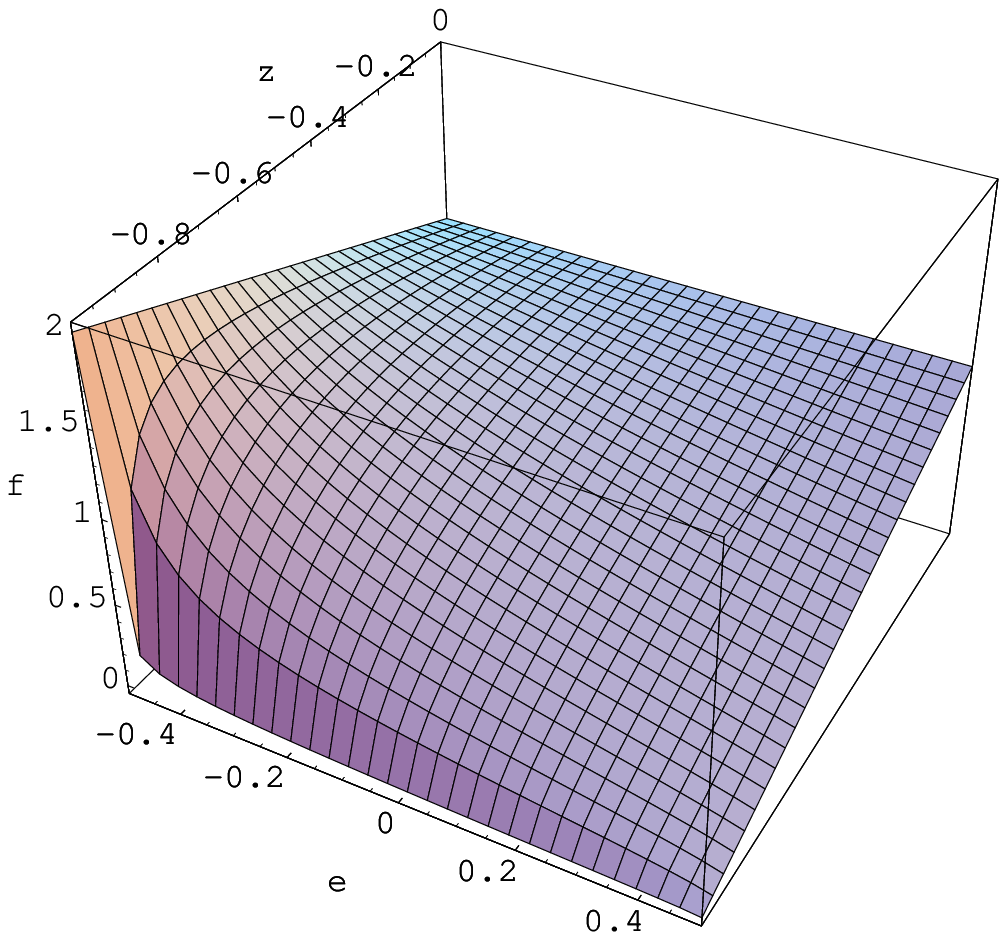}} &
{\input epsf \epsfxsize=5.5cm \epsfbox{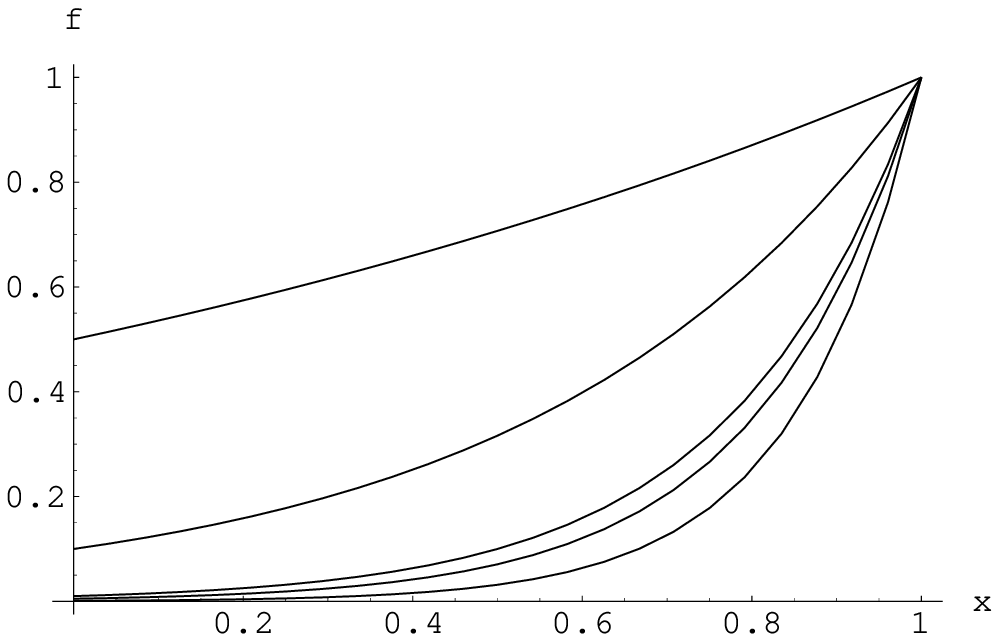}}\\
{} & {}\\
(a) & (b)\\
\end{tabular}
\vs{0.5cm}
{\small Figure 5: a) Plot of $f(\epsilon,z)$
of random "parquet" model in FMSA-like approximation,
b) Plot of $f(x,h)$ in FMSA approximation at $h = 0.5, 0.1,
0.01, 0.005, 10^{-3}.$}
\end{figure}

It follows from fig.4 and fig.5a that all three formulas for $\sigma_{e}$,
despite of their various functional forms,
differ from each other weakly  for
$z \lesssim 0,8$ due to very restrictive boundary conditions (25') and
the exact Keller-Dykhne value. This range of $z$ corresponds approximately
to the ratio $\sigma_2/\sigma_1 \sim 10^{-1}.$ For the smaller ratios a
difference between these functions become distinguishable.

It is useful also to compare $\sigma_e$ in the FMSA approximation with
$\sigma_e^{rm}$ of the Rayleigh model (RM), constructed numerically \ci{14}.
Fig.5b shows $f$ in the FMSA
approximation as a function of $x$ and $h= \frac{1-z}{1+z}$ for the direct
comparison with the numerical results for RM from \ci{14}.
$\sigma_e^{rm}$ satisfies the self-reciprocity relation (17), but does not
satisfy the duality relation (24). For this reason $\sigma_e^{rm}$ at adjoint
concentrations are not related. The critical concentration $x_c$ of the RM
in the percolation limit $z \to 1$ is $x_c = 1-\pi/4 \approx 0.215$ \ci{14}.
It differs strongly from the critical concentration of self-dual systems
$x_c = 1/2.$ By compairing fig.3 from \ci{14} and our figs.4b,5b one can see
that for $z \lesssim 0.8$ both $\sigma_e$ are very similar and
close to each other. For $z \ge 0.8$ they begin to differ, our $\sigma_e$  becomes
smaller than $\sigma_e^{rm}$ for concentrations in the range
$0.2 \le x \le 0.8$ due to a fact, that $\sigma_e$ from (58) must follow the
exact value (22) at $x=1/2.$ At the same time both conductivities remain close
for concentrations in small regions near $x=1$ and $x=0.$
This behaviuor can be explained by the difference in the geometry and symmetry
of both systems, which becomes significant in this range of concentrations.
For this reason one can expect, that for the RM with round inclusions
deformed in a such way that the approximate symmetry (20) appears, its
$\sigma_e^{rm}$ will become more similar to $\sigma_e$ from (58).

\bs
\cl{\bf 8. Percolation limit}
\bs

Now let us consider in the more details the derived formulas for
$\sigma_{e}(\epsilon,z)$
in case when $\sigma_2 \to 0 (z \to 1).$ It is clear that for regularly
inhomogeneous medium one can always construct such distribution of the
conducting phase that  $\sigma_{e}(\epsilon,1)$ will differ from zero for all
$1/2 \ge \epsilon > -1/2$.
But in the case of randomly inhomogeneous medium the limit $\sigma_2 \to 0$
is equivalent to the well known percolation problem [17]. In terms
of $z$ it corresponds to the limit $z \to 1$ and is also similar to the
ideal conductivity limit $\sigma_1 \to \infty$.
Strictly speaking,  an implementation of the duality
transformation (14) is not obvious in this case.
However, if one supposes that the dual symmetry relation (24) fulfills in this
limit too due to a continuity then it follows from (24) that
$$
\sigma_{e}(\epsilon) \sigma_{e} (-\epsilon) = 0.
\en(71)
$$
The relation  (71) does not contradict to the known basic results
of the percolation theory that
$\sigma_{e}(\epsilon) = 0$ for $\epsilon \le 0$ and
$\sigma_{e}(\epsilon) \ne 0$ for $\epsilon > 0$.
Moreover it follows from the general formula (31) that in this case
$\sigma_{s} = |\sigma_a|$ and $\sigma_{e}(\epsilon) = \sigma_a + |\sigma_a|$.
It gives
$$
\sigma_{e}(\epsilon) =
\left\{\begin{array}{rc}
0, & \epsilon \le 0,\\
2\sigma_a, &  \epsilon > 0.\\
\end{array}\right.
\en(72)
$$
This means that a behaviour of  $\sigma_{e}(\epsilon)$ in the
percolation theory is completely determined by its odd part.
From a general discussion of the functional structure of the conductivity
of the two-dimensional randomly inhomogeneous systems and experimental and
numerical results it is known that in the percolation limit the effective
conductivity $\sigma_{e}$ must have a
nonanalytical behaviour near the percolation
edge $x_c = 1/2$ or at small $\epsilon > 0$ \ci{17}
$$
\sigma_{e}(\epsilon) \sim \sigma_1 (x-x_c)^{t} \sim \sigma_1 \epsilon^t,
\en(73)
$$
where a critical exponent of the conductivity $t$ is slightly above 1
and can be represented in the form $t=1+\delta.$
Since the  values of this exponent found by the numerical calculations
are confined to be in the interval (1,10 -- 1,4) [17], then
$\delta$ have to be small and belongs to the interval (0.1 -- 0.4).
It follows from general formula (72) that one must have
$$
f_a(\epsilon, z) \xrightarrow[z \to 1, \epsilon \to 0_+]{}
\epsilon^{t}.
\en(74)
$$
It means that the function $\Phi(\epsilon,z)$ from (30) has at small
$\epsilon$ some crossover on $z$ under $z \to 1$ from a regular (analytical)
behaviour to a singular, powerlike one. At the moment an exact form of
this crossover is unknown. For example, it can be of the form
$$
\Phi(\epsilon,z) \sim
\left(\Phi_0(z^2) + \Phi_1(z^2) \epsilon^2\right)^{\delta(z)/2},
\quad z \to 1,
\en(75)
$$
where $\Phi_0(z^2) \to 0,$ $\Phi_1(z^2) \to \Phi_1 \ne 0$ and
$\delta(z) \to \delta \ne 0$ when $z \to 1.$  Using (75) together with (30)
and (32) one can
show that this crossover takes place at $1-z \sim \epsilon^{2t}$ in agreement
with result obtained from the similarity hypotheses \ci{19,20}.

It follows from the formulas for $\sigma_{e}$ obtained in the sections 4-6,
that one gets always  $\sigma_{e} \to 0$ in the limit  $\sigma_2 \to 0$,
except the region near $x=1.$
It means that these formulas obtained in the FMSA approximation
are not valid in the limit $\sigma_2 \to 0.$
This is confirmed by the appearance of the divergencies
in the expansions of $\sigma_{e}$ in small concentrations in the limit
$z \to 1.$
This fact is a consequence of the assumptions made in the used approximations.
For example, in case of the model of the layered (or striped) squares, where
a deviation from the percolation type behaviour in the limit $z \to 1$ is
more obvious,
this is due to the "closing" (or "locking") effect of the layered  structure
in the adopted approximation in the limit $\sigma_2 \to 0.$ In order
to obtain a finite conductivity in this model above threshold concentration
$x_c$ one needs to take into account the correlations between adjacent squares.
It is easy to show that near the threshold an effective conductivity is
determined by random conducting clusters formed out of the crossing
random layers from neighboring elementary squares. As is well known,
the mean size of these clusters
diverges near the percolation threshold \ci{17}
and for this reason the FMSA type approximation cannot be
applicable for the description of $\sigma_{e}$ in the limit
$z\to 1 \, (\epsilon > 0)$ and of
the percolation problem. The divergencies of the expressions in
the percolation limit mean strong changes of these expressions, which must
result in a formation of the (widely believed) universal critical behaviour
(73).
It follows from figs.4,5 that the EM approximation
overestimates $\sigma_{e}$, and both other formulas underestimate it
in the region $z \to 1, \epsilon > 0.$ We hope to investigate this limit in
detail in other papers.

\bs
\cl{\bf 9. Conclusions}
\bs

Thus we have studied the possible functional forms of the effective
conductivity of the two-phase SD system at arbitrary values of concentrations.
A new approximate functional equation for $\sigma_e$, applicable for
inhomogeneous self-dual systems and generalizing the duality relation,
was deduced in the FMSA approximation and its solution was found.
We have constructed  also a hierarchical ("parquet") model of the random
inhomogeneous stripe type medium and have found its effective conductivity
in the similar approximation at arbitrary phase concentrations. All
expressions for the effective conductivity have the different functional forms.
They

(1) satisfy the self-dual symmetry and all necessary
inequalities,

(2) reproduce the general formulas for $\sigma_{e}$ in the weakly
inhomogeneous case.

All these results confirm a conjecture that, in general, $\sigma_{e}$ already
of two-phase self-dual systems may be a nonuniversal function and
can depend on some details of the structure of the randomly inhomogeneous
regions \ci{21}.
Analogous conclusions for SD systems with $N = 3$ phases were done earlier
in the paper \ci{22}, where
a possibility to find a generalization of the exact Keller-Dykhne
formula for case $N = 3$ was investigated numerically.

The obtained explicit formulas can be used for an approximate description
of the effective conductivity of some real (as regular as random)
inhomogeneous systems like a corresponding formula of the EM approximation.
Another important property of these formulas is connected with a fact that
they admit a generalization on a case of heterophase systems with arbitrary
number of phases $N$ (see \ci{23} and my forthcoming paper).
At the same time the formulas (58),(64) conserve their simple form for
arbitrary $N$, while the EMA
expression for $\sigma_e$ becomes very complicated function even for $N=3.$

\bs
\cl{\bf 10. Acknowledgments}
\bs
The author is especially thankful to V.Marikhin for useful information on a
problem and to Prof. F.Kusmartsev for fruitful discussions and a warm
hospitality at Loughborough University, England.
The discussions with many other colleagues were also fruitful.
This work was supported by RFBR grants \# 2044.2003.2 and \# 02-02-16403.

\end{document}